\documentclass[12pt]{article}
\usepackage{amsfonts}
\usepackage{amssymb}
\usepackage{graphicx}
\usepackage{amsmath}%

\newcommand{\RR}{\mathrm{RR}}

\usepackage{authblk}
\usepackage{natbib}

 \title{On the Flatland paradox and   limiting arguments}
\author{P. Druilhet$^*$\\
$^{*}$Laboratoire de Math\'{e}matiques, UMR CNRS 6620,
Clermont Universit\'e,\\
Universit\'{e} Blaise Pascal, 63177 AUBIERE Cedex, France\\
pdruilhet@univ-bpclermont.fr
}

\begin{document}
\maketitle

\begin{abstract}
We revisit  the flatland paradox proposed by  \cite{ston1976} which is an example of non-conglomerability.   The main novelty in the analysis of  the paradox is to  consider  marginal vs conditional models rather than proper vs improper priors. We show that in the first model a prior distribution should be considered as a probability measure whereas, in the second one, a prior distribution should be considered in the projective space of measure. This induce two different kinds of limiting arguments which are useful to understand the paradox. 
 We also show that  the  choice of a flat prior is not adapted to the structure of the parameter space and we consider  an improper prior based on reference priors with nuisance parameters for which  the Bayesian analysis matches the intuitive  reasoning.

~


 
 \end{abstract}
 
\section{Introduction}

Improper priors are commonly used in Bayesian statistics, especially when no prior information is available. However, using improper priors may lead to some inconsistencies between an intuitive or classical approach and a  Bayesian analysis. The Flatland paradox, introduced by \citet{ston1976}, is an example of such inconsistency and has been largely commented in the literature. The main argument involved to explain the paradox is the improperness of the flat  prior  which leads to non-conglomerability, see e.g.  \citet{scseka1984}, \citet{heathsudd1989}, \citet{jayn2003}. It is also an example of  inconsistency of the limit behaviour of  sequences of proper priors, such as uniform priors with large range.

~

 The aim of this paper is to propose a new way to analyse paradoxes based on the use of improper priors.  Rather than considering  proper vs improper priors, we prefer to consider two different Bayesian paradigms, one associated to the marginal model  and the other one to the conditional model. The way to consider a prior distribution and limiting arguments is quite different from one paradigm to the other and can explain the paradox.  In Section \ref{section.presentation}, we recall the statistical problem. In Section \ref{section.2D}, we show that  the choice of the flat prior is not adapted to the problem in the sense that the parameter of interest used in the intuitive reasoning is not the whole parameter  but a sub-parameter for which the related prior distribution is not flat but highly informative. Then,  we propose an  improper prior that makes a distinction  between  nuisance parameters and parameters of interest and   which corresponds to  the intuitive reasoning. In Section \ref{subsubsection.properprior}, we propose an analysis of the paradox by  limiting arguments based on two different paradigms. We replace the flat prior by a sequence of proper priors and we examine the limit when the range tends to the whole parameter space. 
   In  the first paradigm, there is no inconsistency whereas for the second one, the inconsistency remains even with proper priors, provided that  we reconsider the interpretation of prior distributions and non-conglomerability.

 \section{The Flatland paradox}
 \label{section.presentation}
We give here a presentation of the model as presented in  \citet{stone1982}. Consider  a tetrahedral die which is tossed a unknown  number of times, say $N$ which is probably large. The faces of the die are labelled "a", "b", "$a^{-1}$" and "$b^{-1}$". At each toss, the outcome is recorded subject to the rule that if the outcome is the inverse of the previous one, the two outcomes are removed, i.e. they annihilate each other. So, at the end, we get a path, denoted by $\theta$  with no consecutive  inverse symbols.
 However, $\theta$ is not observed but a supplementary toss is performed and the resulting path, denoted by $x$ is registered following the same rule.
 Observing $x$, a  statistician has to guess $\theta$.
 
 Let $\Theta$ be the set of such finite paths.  We denote  by $x^-$ the  path  obtained from $x$ after removing the last outcome and   by $A_x^+$ the set of the three possible paths obtained from $x$ by adding a symbol without annihilation. Denote  $A_x=A^+_x\cup \{x^-\}$ the four possible paths obtained from $x$. In the special case where $x$ is the null path, denoted by $0$,  $A_0=A_0^+$ is the set of the four possible paths of length $1$ and $x^-$ is not defined. Similarly, we define $\theta^-$, $A_{\theta}^+$ and $A_\theta$.
 For example, if $\theta$ is the path $\ldots abaa$, then $ A^+_\theta=\{\ldots abaaa,\ldots abaab, \ldots abaab^{-1}\}$ and $  \theta^-=\ldots aba$. The likelihood of the model is
 \begin{equation}\label{eq.likelihhod}
 l(\theta;x)=p(x|\theta)=\frac{1}{4}\mathbf{1}_{\theta\in A_x}=\frac{1}{4}\mathbf{1}_{x\in A_\theta},
 \end{equation}
 where $p(x|\theta)=\mathbb P(X=x|\theta)$.
 Given any non-null $\theta$, the event "there is no annihilation" (at the last toss) can be written "$x\in A_\theta^+ $" and we have
 \begin{equation}
 \label{eq.probaAplustheta}
 \mathbb P(\textrm{"no annihilation"}|\theta)=  \mathbb P(X\in A^+_\theta\,|\,\theta)=
 \left\{
 \begin{array}{cl}
 \frac{3}{4}&\textrm{if } \theta\neq 0, \\
 1&\textrm{if } \theta=0. \\
 \end{array}
 \right.
 \end{equation}
 
 So, with probability greater or equal to $3/4$, the path $x$ will be longer than the path $\theta$ for any non-null path    $\theta$. Intuitively, a good estimate of $\theta$ is $\hat\theta=x^-$, the only $\theta$ for  which there is no annihilation.

 This statistical model was first  proposed by \citet{lehm1959} to give an example of a best equivariant estimator which is not admissible. Identifying $\Theta$ as the free group generated  by $a$ and $b$, then the minimax equivariant estimator for $\theta$ under the 0/1 loss function is any  $\widetilde\theta $  such that $\widetilde\theta(x)$ belongs to $A_x^+$. However, the intuitive estimator $\hat{\theta}=x^-$ dominate uniformly $\widetilde{\theta}$ for the associated risk function.

  \citet{ston1976} propose a Bayesian version of this inferential problem and put    a  flat prior $\pi(\theta)\propto 1$   on   $\theta$, which corresponds to the right Haar measure of the free group which is known to be associated to the best equivariant estimator. The posterior distribution is therefore:
 
 $$
 \pi(\theta|x)\propto l(\theta;x)\;\pi(\theta)= \frac{1}{4} \mathbf{1}_{\theta\in A_x}.
 $$
 Given a non null path $x$, the event "no annihilation" can be written "$\theta= x^-$" or $"x\in A^+_\theta"$ with a  posterior probability
 
 \begin{equation}
 \label{eq.probaAplusx}
 \mathbb P (\textrm{"no annihilation"}|x)=\left\{
 \begin{array}{cl}
 \frac{1}{4}&\textrm{if } x\neq 0, \\
 0&\textrm{if } x=0. \\
 \end{array}
 \right.
 \end{equation}
 
 The   inconsistency between (\ref{eq.probaAplustheta}) and (\ref{eq.probaAplusx}), named "the Flatland paradox" after \cite{ston1976},  is an example of non-conglomerability \citep{defi1972,kada1986}. We have simultaneously  $ \mathbb P("\textrm{no annihiliation}"\,|\,\theta)\geq \frac{3}{4}$ for any $\theta$ and  $ \mathbb P("\textrm{no annihiliation}"\,|\,x)$ $\leq \frac{1}{4}$ for any $x$. More generally  a non-conglomerability phenomenon occurs for some event, say $A$, if there exists  $0<a<1$ such that
 \begin{eqnarray}
 \label{eq.nonconglox}
 &\mathbb P(A|x)<a,&\forall x\\
\label{eq.nonconglotheta} \textrm{and}&\mathbb P(A|\theta)>a,&\forall \theta.
 \end{eqnarray}

  Of course, if the prior distribution  $\pi$ were a probability distribution, such inconsistency could not  occur since 
  $$
  \int\mathbb P(A|x) p(x)\;dx=\int\mathbb P(A|\theta) \pi(\theta)\; d\theta,
  $$
 where $p(x)=\int p(x|\theta)\pi(\theta)\,d\theta$ is the marginal probability distribution  of $x$. Note that if $\pi$ is improper, then $p(x)$ is no longer a probability distribution. Other examples of  non-conglomerability and their analysis by using finitely additive probability can be found in \citet{kada1986}.

   In the following, we   give new insights of this phenomenon.  
 We denote by $\ell(\theta)$ the length of $\theta$, by $\ell(x)$ the length of $x$ and  by  $n_\ell$ the number of paths of length $\ell$.
We have $
n_0=1$, $n_1=4$, and $  n_\ell=4\times 3^{\ell-1},$ $  \ell\geq 2.
$ 

 The importance of examining the distribution of $\ell(\theta)$ has been first pointed out by \citet{Hill1979} and will be the key point for one part of the explanation of the paradox.

\section{A  2-dimensional re-parameterization approach}
\label{section.2D}

In this section, we show that a flat prior on $\theta$ induces a highly informative prior distribution on $\ell(\theta)$. This prior does not correspond to  the intuitive approach that suggests a flat prior on $\ell(\theta)$.  Therefore, we   propose another improper prior on $\theta$ that consider  $\ell(\theta)$ as a parameter of interest and the specific path of a given length as a nuisance parameter. For this prior, the paradox disappears.

To explore the features of an improper prior $\pi$, we   define  risk ratios, or a relative weights, of any two finite events A and B:
$$
\RR(A;B)=\frac{\pi(A)}{\pi(B)}.
$$
It is worth noting that  $\RR(A;B)$ does not depend on the arbitrary chosen scalar factor in the definition of $\pi$.
The improper prior distribution on $\ell(\theta)$ derived from the flat prior on $\theta$ is, for  $\ell\geq 1$,
\begin{equation}
\label{eq. priorell}
\pi(\ell(\theta)=\ell)\propto4\times 3^{\ell-1}\propto 3^{\ell},
\end{equation}
and therefore, for $k\geq 2$, we can defined a risk ratio
\begin{equation}
\label{eq.RRl}
\RR(\ell(\theta)=k+1;\ell(\theta)=k-1)= \frac{\pi(\ell(\theta)=k+1)}{\pi(\ell(\theta)=k-1)}=\frac{n_{k+1}}{n_{k-1}}=9.
\end{equation}
So, there is 9 times more "chance" that $\ell(\theta)$ is equal to $k+1$ rather than $k-1$, or, from another point of view, the prior puts   a weight $9$ times larger
on  $k+1$   than on $k-1$ for $\ell$.

Let rewrite   (\ref{eq.probaAplustheta}) and (\ref{eq.probaAplusx}) respectively as:
\begin{equation}
\label{eq.ratiogiventheta}
\mathrm{ODD}(\textrm{"no annihilation"}|\theta)=\frac{\mathbb P(\textrm{"no annihilation"}|\theta)}{\mathbb P(\textrm{"annihilation"}|\theta)}=\frac{\mathbb P(X\in A^+_\theta|\theta)}{\mathbb P(X= \theta^-|\theta)}=3
\end{equation}
and
\begin{equation}
\label{eq.ratiogivenx}
\mathrm{ODD}(\textrm{"no annihilation"}|x)=  \frac{\mathbb P(\textrm{"no annihilation"}|x)}{\mathbb P(\textrm{"annihilation"}|x)}=\frac{ \pi(\theta=x^-|x)}{\pi(\theta\in A^+_x|x)}=\frac{1}{3}.
\end{equation}
Note that $\mathbb P(\cdot)$ refer to the probability of events that involve both $X$ and $\theta$. So, the inconsistency between  (\ref{eq.probaAplustheta}) and (\ref{eq.probaAplusx}) corresponds to the  factor $9$ between  (\ref{eq.ratiogiventheta}) and (\ref{eq.ratiogivenx}). To see that this factor comes from (\ref{eq.RRl}),
let restate the   reasoning in term of $\ell$. For $\ell(\theta)\geq 1$, the probability  that, for a given $\theta$,  there will be no annihilation can be written
$$
\mathbb P(\ell(X)=\ell(\theta)+1)\;|\;\theta)=\frac{3}{4}.
$$
Since this expression depends on $\theta$ only through $\ell(\theta)$, we have
$$
\mathbb P(\ell(X)=\ell(\theta)+1) \;|\;\ell(\theta) )=\frac{3}{4}
$$
or, equivalently, for $k\geq 1$,
$$
\frac{\mathbb P(\ell(X)=k+1 \;|\;\ell(\theta)=k)}{\mathbb P(\ell(X)=k-1 \;|\;\ell(\theta)=k) }=3.
$$
Now, the posterior relative risk of no annihilation vs annihilation is
$$
\frac{\pi(\ell(\theta)=k-1|\ell(x)=k)}{\pi(\ell(\theta)=k+1|\ell(x)=k)}=
\frac{\mathbb P(\ell(x)=k \;|\;\ell(\theta)=k-1 )}{\mathbb P(\ell(x)=k  \;|\;\ell(\theta)=k+1 ) }\times \frac{\pi(\ell(\theta)=k-1)}{\pi(\ell(\theta)=k+1)}= \frac{3}{9}=\frac{1}{3}.
$$

It can be seen how the  prior distribution involved in this  formula influences the posterior distribution. So, the inconsistency between the intuitive and Bayesian solutions comes from  the fact that the prior is highly informative on the parameter of interest $\ell(\theta)$. 
To show that the paradox is not directly related to the improperness of the prior but to its construction, we   propose another   improper prior, say $\tilde \pi$,  which is  implicitly used is the intuitive reasoning and for which the inconsistency disappears.

We do not  know  $\ell(\theta)$ but we just think that   $\ell(\theta)$ is probably large   and  we implicitly assume that, as a prior knowledge,  the event $"\ell(\theta)=k"$ is almost as likely as the event $"\ell(\theta)=k-1"$ or that $"\ell(\theta)=k+1"$ especially for large values of  $k$, so we put,  as an approximation, a flat prior on $\ell(\theta)$. Knowing the length $\ell$ of $\theta$ and using symmetry arguments,  any path of length $\ell$ has the same probability to be drawn, which is equal to $\frac{1}{4\times 3^{\ell-1}}$ for $\ell\geq 1$.
The resulting prior on $\theta$ is, for $\ell(\theta)\geq 1$,
\begin{equation}\label{eq.prior2}
\widetilde \pi(\theta)\propto\widetilde \pi(\theta|\ell(\theta))\;\widetilde \pi(\ell(\theta))\propto \frac{1}{4\times 3^{\ell(\theta)-1}} \propto \frac{1}{3^{\ell(\theta)}}
\end{equation}

   The prior $\tilde \pi$ can also be obtained by using   reference priors with nuisance parameters as  proposed by  \citet{bern1979,bernberg1992} or \citet{kaswas1996}: the parameter $\theta$ can be split  into two parameters:  $\theta=(\ell,\eta)$, where $\ell$ is the length of the path and $\eta=1,..,n_\ell$ is the index of the path within the pathes of length $\ell$. We can consider $\ell$ as the  parameter of interest and $\eta$ as a nuisance parameter.    The reference prior on $\ell$ is the flat prior and  we know exactly the distribution of $\eta|\ell$ which is a uniform distribution over $\{1,2,...,n_\ell)$. The resulting prior is therefore $\widetilde \pi$.

For $\ell(x)\geq 2$, the posterior distribution is
$$
\widetilde \pi(\theta|x)\propto \frac{1}{ 3^{\ell(\theta) }} \mathbf{1}_{\theta\in A_x}.
$$
and therefore
\begin{equation}
\frac{\mathbb P(\textrm{"no annihilation"}|x)}{\mathbb P(\textrm{"annihilation"}|x)}=\frac{ \widetilde \pi(\theta=x^-|x)}{\widetilde \pi(\theta\in A^+_x|x)}=\frac{3^{\ell(x)+1}}{3\times 3^{l(x)-1}}=3.
\end{equation}
which matches  (\ref{eq.ratiogiventheta}) or equivalently
$\mathbb P(\textrm{"no annihilation"}|x)=3/4$ which matches (\ref{eq.probaAplustheta}).  
So,  the intuitive reasoning is in fact a Bayesian reasoning  using the improper distribution (\ref{eq.prior2}) for which the paradox disappears.  We may note that a similar effect is seen in the marginalization paradox by \citet[Example 1]{stda1972} where the paradox disappears when using another improper prior. 

 ~

\noindent\textit{Remark}:  \citet[p. 453]{jayn2003} tried to explain the paradox by exploring the link between the  priors on $N$, the number of tosses  and the priors on $\ell(\theta)$. We think that this leads to unnecessary complications   since the problem  can be restated  as follows : from the second toss,  instead of drawing at random one of the four faces of the tetrahedral die, we draw at random with equal probability one of the three letters that is not the inverse of that appearing in the previous outcome. Only at the supplementary toss that generate $x$,  one letter among the four possible  ones is drawn at random, which may lead to a  possible  annihilation. In that case, $\ell(\theta)=N$   and the paradox, i. e. the inconsistency  between (\ref{eq.probaAplustheta}) and (\ref{eq.probaAplusx}) remains the same.

\section{An approach by  a limiting argument}
\label{subsubsection.properprior}

In this section, we propose another approach of the problem by considering limits of proper priors.
Let   replace the flat prior $\pi$  by  $\pi_M $, the uniform prior on the paths of lengths lower or equal to $M$. When $M$ is large, it is commonly admitted that $\pi_M$ is an approximation of  the flat prior.
 Since the number of paths with lengths between $0$ and $M$ is equal to $2\times 3^M-1$, $\pi_M(\theta)$ is defined by
\begin{equation}
\label{eq.uniformproper}
\pi_M(\theta)=\frac{1}{2\times 3^M-1}\mathbf 1_{0\leq\ell(\theta)\leq M}.
\end{equation}
Equivalently, $\pi_M$ can be seen as a result of a two steps random procedure corresponding to the parameterization $(\ell,\eta)$ described in Section \ref{section.2D}:
\\
 - step 1: draw at random a length $\ell$, according to the distribution $\pi_M(\ell)=\frac{4\times 3^{\ell -1}}{2\times 3^M-1}$ if $\ell\neq0$ and $\pi_M(0)=\frac{1}{2 \times 3^M-1}$.
 \\
 - step 2: draw at random a path  $\theta$ of length $\ell$ from a uniform prior:  $\pi(\theta|\ell)=\frac{1}{4\times 3^{\ell-1}}\mathbf{1}_{\ell(\theta)=\ell}$ if $\ell\neq0$. If $\ell=0$, $\theta$ is the null path with probability 1.

To understand the notion of approximation and its implication when $M$ goes to $+\infty$, it is necessary  to distinguish between two Bayesian paradigms corresponding respectively to a subjective and an objective approach. 
 
~

\emph{Paradigm 1 (subjective approach)}: we assume  that $\theta$ is drawn at random according to $\pi_M$. Therefore, $\theta$ can be considered as a random effect with known distribution $\pi_M$. We may note that an improper prior $\pi$ is not relevant in this approach since it is not a probability distribution. The relevant model for $x$ is the marginal model $p_M(x)=\sum_{\theta} l(\theta;x)\pi_M(\theta)$. Changing  $
M$ implies that the way  $x$ is generated  also  changes, which means that it is irrelevant to consider the  limit of the  posterior distribution with respect to $\pi_M$  for $x$ fixed. Therefore, in the limiting argument, it is essential to consider the behaviour of $x$.

Since $\pi_M$ is a probability, we have:
\begin{eqnarray}
\label{probajointe}
\mathbb P_M("\textrm{no annihiliation}")&=&\sum_\theta\pi_M(\theta) \; \mathbb P("\textrm{no annihilation}"\,|\,\theta)\\
\label{probajointe2}
&=&\frac{3}{4}+\frac{1}{4}\pi_M(0) \approx \frac{3}{4},
\end{eqnarray}
where $\pi_M(0)$ is negligible  for $M$ large.  Clearly, we also have 
\begin{eqnarray}
\label{probajointe}
\mathbb P_M("\textrm{no annihiliation}" )
&=&\sum_x p_M(x) \; \mathbb P_M("\textrm{no annihilation}"\,|\,x) 
\end{eqnarray}

From (\ref{probajointe2}) and \ref{probajointe},  on average over $x$, $\mathbb P("\textrm{no annihiliation}"|x) $ is almost equal to 3/4, which corresponds to the intuitive reasoning and the standard probability rules.
 Knowing $x$ gives more information:
 \begin{equation}
  \label{eq.probapiM cond}
\mathbb P_M(\textrm{"no annihilation"}|x)=
\left\{
  \begin{array}{ccc}
    1  &\mathrm{if}&  \ell(x)= M  \textrm{ or } M+1,\\
    \frac{1}{4}&\mathrm{if}& 1\leq\ell(x)\leq M-1 ,\\
   0&\mathrm{if}& \ell(x)=0 .\\
  \end{array}
\right.
  \end{equation}
By a straightforward calculation,  the event 
$
\mathbb P_M\left(l(X)=M\textrm{ or }  M+1\right)\approx \frac{2}{3}
$
and
 $\mathbb P_M(1\leq\ell(X)<  M  ) \approx \frac{1}{3}$.
 So, for the prior $\pi_M$, the difference between (\ref{eq.probaAplustheta}) and (\ref{eq.probaAplusx}), which was called inconsistency for the flat prior, remains, but only when $\ell(x)<M$ which occurs with  probability $\approx 1/3$.
  This can be explained intuitively, as in the improper case, by the fact  that there is $9$ times  more chances that $\ell(\theta)=\ell(x)+1$ rather than $\ell(\theta)=\ell(x)-1$ when $1\leq \ell(x)\leq M-1$.   So, given $x$,  it is more likely  that   $\ell(\theta)=\ell(x)+1$ with annihilation than  $\ell(\theta)=\ell(x)-1$ without  annihilation.

Now, let $M$ go to $+\infty$. As mentioned above, considering the limit of
$\mathbb P_M("\textrm{no annihiliation}"|x)$ or more generally $\pi_M(\cdot|x)$  for $x$ fixed  is not  relevant: if $M$ get larger, the values of $\ell(x)$  get larger with a high probability.  For example, it is not possible to  replace $\pi_M$ by $\pi$ in  (\ref{probajointe}) for the limiting expression since it gives   $+\infty$ whereas the limit is $3/4$. In Eq. (\ref{probajointe2}), $p(x)$ is formally equal to $1$ for the flat prior, and therefore is not defined as a probability distribution on $x$ \citep[see][]{taralind2010}.
   This illustrates the fact that the flat prior cannot be considered as the limit case of inference with $\pi_M$ and that limiting arguments are not valid.
 So, when $M$ goes to $+\infty$, we can only  say that $\ell(x)$ goes to $+\infty$ in the sense that   $\mathbb P_M(\ell(X)\leq k)$ goes to $0$ for any fixed $k$.  The most probable case, that is $"l(X)=M\textrm{ or }  M+1"$, which corresponds to "no annihilation" with conditional probability 1, varies with the prior, as pointed out by \citet{stone1982}.

~

\emph{Paradigm 2 (objective approach):} we consider that there is some $\theta$ such that $x$ has been generated according  to $p(x|\theta)$. The relevant model for $X$ in this  paradigm  is  the conditional model $p(x|\theta)=l(\theta;x)$ rather than the marginal model $p_M(x)$  in the subjective approach. Here,  $\theta$ is considered as a fixed parameter  rather than a random effect. In that case,   $\pi_M$ is not the actual way to generate $\theta$, but a way to make inference on $\theta$. To cite  J.M.  Bernardo in \cite{ironbern1997}, "one should not interpret any non-subjective prior as a probability distribution". So, it  is   not relevant in this paradigm to give an interpretation of the marginal distribution  $p_M(x)$ of $X$ neither of
the joint distribution of $(X,\theta)$ based on $\pi_M$. Therefore, it is irrelevant to consider $\mathbb P_M("\textrm{no annihiliation}")$ in (\ref{probajointe}) which is related to the joint distribution.
 Changing the prior distribution will not change the way to generate $x$, but will only change  the posterior distribution and the related inference on $\theta$. It is therefore relevant here  to consider the limit of the posterior distribution  for  $x$ fixed. Moreover, for any scalar $\alpha>0$, $\pi$ and $\alpha \pi$ give the same posterior distribution which means that prior distributions are defined up to a scalar factor.  This leads to consider prior distributions in a projective space of measures and not as probabilities  (see \citet{bidr2016} or \citet{taralind2016}).

 In the projective space, improper priors  appear  naturally as limit of proper prior sequences for the  corresponding convergence mode, named $q$-vague convergence \citep{bidr2016}: 
a  sequence $\{\pi_M\}_{M\in\mathbb N}$  of discrete priors   is said to converge $q$-vaguely to the discrete prior $\pi$ if there exists some scalars $a_M$ such that $a_M \pi_M(\theta)$ converges to $\pi(\theta)$ for any $\theta$. Here, choosing $a_M=2\times 3^M-1$, it is easy to see that  $\pi_M$ converges to the flat prior $\pi$. Therefore, contrary to Paradigm 1,
 $\pi_M$ can  be seen as an approximation of the flat prior.   Note that the $q$-vague convergence of $\pi_M$ implies the convergence for the posterior distribution for $x$ fixed.   For example, from (\ref{eq.probapiM cond}),  $\lim_{M\rightarrow + \infty}\mathbb P_M(\textrm{"no annihilation"}|x)=\frac{1}{4}$   which matches with $\mathbb P(\textrm{"no annihilation"}|x)$ under  the flat prior. This illustrates the fact  that   limiting arguments for $x$ fixed  are valid in the second paradigm.

We now justify the fact that,  in the second paradigm, the improperness of the prior is not directly involved in the inconsistency  by showing that the inconsistency remains with proper priors sufficiently close to the flat prior.
 In practice,  a statistician has a vague idea of the range in which $\theta$ lies and instead of choosing a flat prior, he will choose a prior of type $\pi_M$. If he thinks  that the length of $\theta$ is probably not greater than some hundred of thousands, he will choose $M$ equal to some millions in order to be sure to encompass the true value of $\theta$  with a sufficient  faith. He  assumes implicitly that the precise choice of $M$ will not have a great influence on the results since it does not assume that the prior  represents the actual way to draw $\theta$. This is the case here, since    provided that $\ell(\theta)<M-1$, the posterior distribution $\pi_M(\theta|x)$ does not depend on $M$.   So, if the statistician is almost certain that $\ell(\theta)<M-1$, then he is also almost certain that $\ell(x)<M$. So, for those $x$,  the inconsistency between the  intuitive and the  Bayesian answers remains. The non-conglomerability is also achieved by a proper prior if we change the  condition "$\forall x$" in (\ref{eq.nonconglox}) by the condition "for any expected $x$". We see again than  the inconsistency is due to the inappropriate choice of the flat prior as prior knowledge, as explained in Section \ref{section.2D} rather than its improperness.

\section{Discussion}

  The Flatland paradox is a striking example where a flat prior or a right Haar measure on a discrete parameter cannot be considered as a non-informative prior or as a reflect of ignorance. This is mainly due to the structure of the parameter space. As described by \cite{abbo18}, changing the dimension of a space may change the  shape of its subsets and mountains may appear flat.
  This is the case here: the parameter space $\Theta$ can be defined in one dimension
  $\Theta=\{\theta_n,n\in\mathbb N\}$ or in two dimensions
  $\Theta=\{(\ell,\eta)\,;\, 1\leq \eta\leq n_\ell, \ell \in \mathbb N\}$. A flat prior on $\theta$ gives an exponentially increasing prior for $\ell$.
  The fact that a flat prior on $\ell$ gives a satisfactory answer   shows that  there does not exist an automated way to choose a prior when no information is available or when we want to ignore the partial information we have.

 The Flatland paradox, as many other  paradoxes or inconsistencies that arise in Bayesian inference with improper priors,   suggests that there is a gap between proper and improper priors.  On the other side, improper priors are often considered as limits of proper priors. 
 Rather than considering  proper vs improper priors, we prefer to  make a distinction between   two different Bayesian paradigms. Each paradigm has its own rules and mixing the rules from one paradigm to the other one generates paradoxes. This is the case for limiting arguments that are quite different from one paradigm to the other one.
 
 In the first paradigm,  $\theta$ is considered as a random effect and the prior distribution should be considered as a way to draw the parameter. The prior distribution must be  a probability distribution and improper priors should be excluded. The relevant model for the data is the marginal model and  non-conglomerability phenomenon cannot occur according to standard probability rules. Limiting arguments with respect to the prior distribution  should include the fact that the marginal distribution of $x$ also changes and improper priors do not appear as limits of proper priors as we have shown.

  In  the second paradigm, $\theta$ is an unknown parameter and the relevant model for $x$  is the conditional model. Prior distributions should be considered in a projective space of measures, i.e. defined up to a scalar factor, rather than a probability distributions. There is no reason in this paradigm to exclude improper priors which appear naturally as  limits of  proper  priors \citep[Theorem 2.6]{bidr2016} independently of the statistical model.  The rules associated to projective spaces are quite different to that associated to probability distributions and non-conglomerability phenomenon may occur. Limiting arguments should be  considered with  $x$ fixed. The convergence mode associated to the projective space is the $q$-vague convergence and can explain  for example, the Lindley paradox \citep{bidr2016}.  
 
 A general consequence of this approach is that inconsistencies in the limit may arise  in  equations involving the joint  or the predictive distributions  when we replace a sequence of    proper priors $\pi_n$ by its   limit $\pi$  in the sense of the q-vague convergence.  Indeed,  joint  or   predictive distributions are associated to the first paradigm whereas the q-vague convergence is associated to the second one. 
 This is the case for the non-conglomerability example analysed in this paper as e.g. in Eq. (\ref{probajointe}). However, the limit involving the posterior distribution with $x$ fixed is consistent as in Eq.  (\ref{eq.probapiM cond}).
 
 This approach suggests a general method to analyse inconsistencies or paradoxes. If the reasoning involves the joint or predictive  distribution, then, only proper priors that reflect a subjective knowledge should be considered and the relevant model is the marginal model. Improper priors are allowed only in an objective approach, where the relevant model is the conditional model and where improper priors can be considered as limits of proper priors.  
   
\subsection*{Acknowledgment}

The author would like to thank the referee for his detailed and helpful comments.

\bibliographystyle{apalike}
\bibliography{flatland}

\end{document}